\documentclass[twocolumn,prb,superscriptaddress,preprintnumbers,amsmath,amssymb]{revtex4}

\usepackage{graphicx}
\usepackage{amsmath,amssymb}
\usepackage{bm}% bold math
\usepackage{ifthen}
\usepackage{dcolumn}% Align table columns on decimal point
\usepackage{times}

\begin{document}

\bibliographystyle{apsrev}

\title{Experimental imaging and atomistic modeling of electron and hole  
quasiparticle wave functions in InAs/GaAs quantum dots.}
\author{G. Bester}
\affiliation{National Renewable Energy Laboratory, Golden CO 80401}
\author{D. Reuter}
\affiliation{Ruhr-Universit\"at,  44801 Bochum, Germany}
\author{Lixin He}
\author{A. Zunger}
\affiliation{National Renewable Energy Laboratory, Golden CO 80401}
\author{P. Kailuweit}
\author{A. D. Wieck}
\affiliation{Ruhr-Universit\"at,  44801 Bochum, Germany}
\author{U. Zeitler}
\author{J. C. Maan}
\affiliation{High Field Magnet Laboratory,  6525 ED Nijmegen, The Netherlands}
\author{O. Wibbelhoff}
\author{A. Lorke}
\affiliation{Universit\"at Duisburg-Essen, 47048 Duisburg, Germany}
\date{\today}

% Abstract

\begin{abstract}
We present experimental magnetotunneling results and atomistic pseudopotential calculations of quasiparticle  
electron and hole wave functions of self-assembled InAs/GaAs quantum dots. The combination of a predictive theory along with the experimental results allows us to gain direct insight into the quantum states. We monitor the effects of (i) correlations, (ii) atomistic symmetry and (iii) piezoelectricity on the confined carriers and (iv) observe a peculiar charging sequence of holes that violates the Aufbau principle.
\end{abstract}

\maketitle

% Introduction
\section{Introduction}
The localized, quantized and entangled states of carriers confined to quantum dots manifest a wealth of novel physical phenomena that are observed experimentally mostly through measurements of the  {\em energies}  of characteristic processes such as the formation, splitting, and charging of excitons \cite{Bimberg99,Woggon97,reimann02,Jacak98}.
However, yet another level of control might be achieved through engineering of the {\em wave functions} themselves through manipulation of their degree of localization, spatial anisotropy or angular-momentum character. The first crucial capability that would allow the design of a device on the fundamental level of its wave function character is wave function monitoring along with an understanding of the controlling physical parameters. However, both experimental imaging \cite{vdovin00,patane02,wibbelhoff05,millo01,venema99,johal02} and theoretical calculations of the many-particle wave functions of dots present a formidable challenge.  From the theoretical point of view, the tradition has largely been to fit a few measured energy levels by adjusting a few parameters in the confining potential within simple models (e.g parabolic effective-mass models), or directly adjusting the energy-related quantities (e.g., tunneling in dot molecules or fine-structure splittings), without imposing  physical reality on the wave functions other than  their  boundary conditions. However, theoretical
 determination of wave functions are more challenging than modeling of the corresponding eigenenergies because of their high sensitivity to subtle physical effects. For instance, piezoelectric terms or particle-particle correlations have a rather small effect on energies\cite{bester06b} yet can affect wave function shapes rather clearly. 
In this contribution, we present a combination of experimental magnetotunneling technique and a predictive theoretical modeling of wave functions that gives us an unprecedented insight into the physics of carriers in confined geometries. This combined approach provides a direct answer to important
physical problems such as the hole filling sequence \cite{reuter05,he05} that violates the Aufbau principle, the importance of the atomistic symmetry, correlations and  piezoelectricity and we address the decade-old question \cite{grundmann95} whether the anisotropy observed in the spectroscopy of self-assembled dots \cite{fricke96} is caused by shape anisotropy (oval dots) or piezoelectric fields. We believe that the type of studies descried here are very promising for wave function engineering and design of future nano-scale devices.

% Experimental Details
\section{Experimental Method}

The systems studied here are self-assembled InAs quantum dots, grown by the Stranskii-Krastanov method\cite{guha90,Bimberg99}. They are investigated by capacitance-voltage (C-V) spectroscopy \cite{drexler94,fricke96,reuter05}. The energy of the quantum dot states are shifted with respect to a carrier reservoir (back contact) by the applied voltage. Characteristic maxima in the capacitance appear, each time an additional electron (or hole) can tunnel into the dots. This makes it possible to determine the addition spectrum of such dots with great accuracy  \cite{miller97}. 

To map out the probability densities of the quantum dot states, we extend the approach of Vdovin \textit{et al.} \cite{vdovin00} and measure the ac-tunneling current between the dots and and  the back contact. 
Additionally, a magnetic field $B$ is applied perpendicularly to the tunneling direction. This field imposes an in-plane momentum
\begin{equation}
\label{eq:kparallel}
k_{||} = \frac{d e B}{\hbar}
\end{equation}
on the tunneling of the carriers, where $d$ is the tunneling distance. At sufficiently high frequencies, the amplitude of a capacitance maximum is a measure of the tunneling rate between the dots and the back contact \cite{luyken99}. The tunneling rate, in turn, is proportional to the probability density of the quasi-particle wave function in k-space \cite{patane02,vdovin00,rontani05}. Experimentally, mapping out the wave functions thus requires recording the C-V amplitudes as a function of the in-plane magnetic field for different azimuthal orientations  \cite{wibbelhoff05}. 

The  investigated samples are GaAs-(Al$_x$Ga$_{1-x}$As) Schottky diodes, grown by molecular beam epitaxy, with embedded InAs quantum dots. 
The p-doped (hole) sample was prepared 
as described in Ref.~\onlinecite{reuter05} however, with a slightly thicker tunneling barrier of 19 nm, to facilitate wave-function mapping.
The layer sequence of the n-doped sample for the electron spectroscopy can be found in Wibbelhoff  \textit{et al.} \cite{wibbelhoff05}. 
Schottky diodes were prepared by alloying ohmic contacts and depositing Cr-Au topgates (300$\mu$m $\times$ 300 $\mu$m). The C-V spectroscopy was carried out using a standard LCR meter (Agilent 4284A) with an ac voltage modulation of $\Delta V = 10$ mV. The frequency was appropriately chosen (8 -- 40 kHz) so that the capacitance amplitude reflects the tunneling rate \cite{luyken99,wibbelhoff05}. 
To determine the tunneling probability as a function of $k_{||}$, and this way map out the quasi-particle wave function in momentum space, C-V spectra for in-plane fields up to $B=26$ T and for azimuthal angles in steps of 15$^\circ$ (starting parallel to [011] crystal direction) were evaluated. The in-plane momentum $k_{||}$ follows from Eq.~(\ref{eq:kparallel}). The normalized C-V amplitudes of the different charging peaks (0 to 6 carriers per dot) are plotted as a function of $k_{||}$ in Figs.  \ref{fig:mts}(a) and (c).

% Theory
\section{Theoretical Method}

A full theory of magneto-tunneling would require a self-consistent calculation of the transport properties of the full system under external field. This is still prohibitive at an atomistic level for such large systems ($\simeq 10^6$ atoms). Instead, we used a simplified transmission theory, which ignores the nonlinear effects of electric field and device structure effects and assumes resonant tunneling, i.e., the emitter state are tuned in such a way to be in resonance with the quasiparticle quantum dot state.
We have calculated the transition rate of an electron or hole from an emitter in state $\kappa$ to a quantum dot containing $N$ particles following the work of Bardeen \cite{bardeen61}.
In this approximation the transition rate is given by $T_{\kappa,N} \propto |\mathcal{M}_{\kappa,N}|^2$ and the transition matrix elements for the transfer of one particle from the emitter in state $\kappa$ to the quantum dot state $|N\rangle$, filled by $N$ electrons, is given by:
\begin{equation}
\mathcal{M}_{\kappa,N} = \int \phi_{\kappa}^*({\bf x}) 
\Psi_{\rm QD}({\bf x}) d{\bf x}\quad .
\label{eq:transm}
\end{equation}
Here, $\phi_{\kappa}^*({\bf r})$ is the probing or emitter wave function and is generated by the external source. $\Psi_{\rm QD}({\bf x})$ is the quasi-particle excitation between the $N-1$ particle states $|N-1\rangle$ and the $N$ particle states $|N\rangle$, i.e.,
\begin{equation}
\Psi_{\rm QD}({\bf x}) 
= \sum_i \langle N-1 | \hat{c}_i |N\rangle\, \psi_i ({\bf x})\quad .
\label{eq:qusiwav}
\end{equation}
$\psi_i ({\bf x})$ is the atomistic $i$th single-particle wave function and $\hat{c}_i$ is an electron (hole) annihilation operator. 

To obtain the correlated many-body states, we use the configuration interaction approach where the many-body wave function is written as a superposition of different Slater determinants (configurations), such as, 
\begin{eqnarray}
\label{eq:CI}
|N-1\rangle &=& \sum_{\alpha} C^{(N-1)}_{\alpha} \Phi_{\alpha}({\bf x}_1, 
\cdots, {\bf x}_{N-1})
 \, ,\nonumber \\
|N\rangle &=& \sum_{\beta}  C^{(N)}_{\beta} \Phi_{\beta}
({\bf x}_1, \cdots, {\bf
 x}_{N-1}, {\bf x}_N) \, ,
\end{eqnarray}
where $\Phi_{\alpha}({\bf x}_1, \cdots, {\bf x}_{N-1})$ is a Slater 
determinant of 
${N-1}$ electrons, and $C_{\alpha}^{(N-1)}$ is its weight. 
Accordingly,  
$\Phi_{\beta}({\bf x}_1, \cdots, {\bf x}_{N})$ is a Slater 
determinant of $N$ electrons, and  $C_{\beta}^{(N)}$ is its weight.
Therefore,
\begin{eqnarray}
\label{eq:mkn}
&&\mathcal{M}_{\kappa,N} = \sum_i \langle N-1 | \hat{c}_i
| N \rangle \langle \phi_{\bf k} | \psi_i\rangle \nonumber \\ 
&&=\sum_i \sum_{\alpha,\beta} C^{(N-1)}_{\alpha}  C^{(N)}_{\beta} 
\langle \Phi^{(N-1)}_{\alpha} | \hat{c}_i | \Phi^{(N)}_{\beta} \rangle\,
\langle \phi_{\kappa} | \psi_i\rangle
\end{eqnarray}
with
\begin{equation}
\langle \Phi^{(N-1)}_{\alpha} | \hat{c}_i | \Phi^{(N)}_{\beta} \rangle
=\left\{ \begin{array}{rl} 
1 &\mbox{if }  |\Phi^{(N-1)}_{\alpha}\rangle= \hat{c}_i | \Phi^{(N)}_{\beta} \rangle,  \\ \nonumber
-1 &\mbox{if }  |\Phi^{(N-1)}_{\alpha}\rangle=- \hat{c}_i 
| \Phi^{(N)}_{\beta} \rangle,  \\
0  & \mbox{otherwise.}\end{array} \right.
\end{equation}
where $i$ is the single particle states. 

In the past, $\mathcal{M}_{\kappa,N}$ was calculated by using single-band effective mass wave functions for $\psi_i$  \cite{rontani05}.
However, an effective mass theory ignores the atomistic character of the wave functions and may miss inter-band and inter-valley effects. Here we use an atomistic empirical pseudopotential approach that takes multi-band, multi-valley and spin-orbit coupling into account.
 
Our atomistic wave functions for the quantum dot states can be written as:
\begin{equation}
\psi_i({\bf x})= \sum_{n}^{N_B} \sum_{\bf k}^{N_k} 
c^{(i)}_{n,{\bf k}}\; u_{n,{\bf k}}({\bf x}) \, e^{i {\bf k}\cdot {\bf x}} \quad ,
\end{equation}
where $u_{n,{\bf k}}({\bf x})$ are the strained InAs bulk Bloch wave functions (linear combination of bulk bands methods \cite{wang99b}), $N_B$ and $N_k$ are the number of bands and  {\bf k}-points, respectively. The probing emitter wave function can be written as:
\begin{equation}
\phi_{\bf k}({\bf x}) = \bar{u}_{\bf k} ({\bf x}) 
e^{i\,{\bf k}\cdot{\bf x}}\, ,
\end{equation}
where $\bar{u}_{\bf k} ({\bf x})$ is the Bloch part of the emitter wave function, and is not precisely known. The projection in Eq.~(\ref{eq:mkn}) can be written as:
\begin{equation}
\langle \phi_{\bf k} |\psi_i\rangle=\sum_n^{N_B} 
\langle \bar{u}_{\bf k}|u_{n,{\bf k}}\rangle
c_{n,{\bf k}}^{(i)} \quad .
\label{eq:bloch-proj}
\end{equation}
Since we do not know the exact form of the Bloch part of the probing
wave function, we assume that $\langle \bar{u}_{\bf k}|u_{n,{\bf k}}\rangle= {\rm const.}$

% GENERAL

\section{Imaging wave functions establishes charging sequence}

The two issues here are: (i) Whereas the {\em single-particle} orbital energies follow the order $S$, $P$, $D$, the addition of carriers may not successively fill the levels in that oder, but skip one shell, violating the Aufbau principle. Furthermore, (ii) the $P$ states may split into $P1$ and $P2$, even if the geometric shape of the dot is perfectly cylindrical. This ``symmetry breaking'' results from the fact that even in perfectly cylindrical/lens-shaped dots made of {\it zincblende} material the atomistic symmetry and the strain symmetry (driving piezoelectricity) is $C_{2v}$ where $P1$ and $P2$ need not be degenerate. Figures~\ref{fig:filling}a) and b) show two different filling sequences for electrons that may result from these issues and that can be distinguished by wave function imaging. Indeed, the fourth electron in Fig.~\ref{fig:filling}(a) tunnels into the $P1$ state while it tunnels into the $P2$ state for filling ``Sequence II''. Since $P1$ and $P2$ have a different geometrical shapes both scenario can be discriminated by mapping out the tunneling amplitudes. 

\section{Effect of the quantum dot shape}
\label{sec:Quantum_Dot_Shape}
One of the challenges posed at the onset of any comparison between theory and experiment is given by the experimental determination of the quantum dot shape, that will subsequently be used in the simulations. According to AFM- and SEM-measurement of the uncapped dots, the shape seems circular. However, it is well known that the shape of the dots changes significantly by the overgrowth process and during this process an anisotropy might be introduced as well as a smoothing of the interfaces by diffusion and exchange between In and Ga. Note that the measurements are performed on a statistical ensemble of dots of slightly different shapes and sizes. 
%Preliminary cross sectional STM measurements \cite{koenraads_private} seem to point out to a 25 nm base diameter and 8 nm height with a strong In composition gradient going from In-rich at the top to In-poor at the bottom of the dots.

For the calculations to be representative and to assess the robustness of our theoretical results, we surveyed a large number of dot shapes (4 different heights, 3 different base sizes and 5 different elongations) and the qualitative features of the results remained. In the present contribution we will show results for three different dots detailed in Table \ref{tab:geometry} and labeled as $D1$, $D2$ and $D3$. $D1$ and $D3$ both have a circular base but different heights and $D2$ has an elliptical base with elongation along the [1$\bar{1}$0] direction. From the magnetotunneling results, we have a hint for an elongation of the dot: In Figure~\ref{fig:elongation} we show contour plots of the transition probability measured by magnetotunneling spectroscopy [(a) and (d)] and calculated for dot $D1$ [(b) and (e)]
and dot $D2$ [(c) and (f)]. We show results for the tunneling of the first electron [(a),(b),(c)] and the first hole [(d),(e),(f)] into quantum dot states with strongly dominant orbital $S$-character. The experimental results show an elongation of the signal along the [1$\bar{1}$0] direction in real space. The theoretical 
results for the elongated dot $D2$ agree well with the experiment while the dot $D1$ with a circular shape differs from the experimental picture. We interpret this result as a strong hint for a structural elongation and conclude that dot $D2$ is the one closest to the experimental reality. 

\section{Addition Energies}

In Table~\ref{tab:addition} we summarize the experimental and calculated addition energies $\Delta(N-1,N)$. We calculated the addition energies from our many-body energies resulting from configuration interaction. We define $\Delta(N-1,N)$ as the difference between the charging energies for $N$ and $(N-1)$ particles, $\mu (N)$ and $\mu (N-1)$ respectively:
\begin{eqnarray}
\Delta(N-1,N) &=& \mu(N) - \mu (N-1)\\
&=& E(N) - 2E(N-1) + E(N-2)	\quad .
\end{eqnarray}
The charging energy $\mu (N)$ is the energy required to add an additional carrier to a dot already occupied by $N$ charges. 

Both electron and hole addition energies in Table~\ref{tab:addition} agree fairly well with the experimental results. As a general trend, the addition energies for the holes are underestimated by the theory while no such general underestimation can be observed for the electron addition energies. The difference in the bare magnitude of the addition energies can therefore not be attributed to a simple difference in size between the experimental and the simulated dots. 

The failure of the addition energies to unambiguously identify the best choice between dot $D1$, $D2$ and $D3$ (table~\ref{tab:geometry}) highlights how complementary the magnetotunneling results are in pinpointing salient features of the quantum dots, such as elongation not revealed in charging or addition energies.

% ELECTRONS
\section{Results for Electrons}

Figure \ref{fig:mts}(a) shows the experimentally determined quasi-particle wave-functions of electrons for different occupation numbers $N$.  
The wave functions of the lowest two states ($N$ = 1 and $N$ = 2) are clearly $S$-like and their similarity suggests weak correlations in the $N$ = 2 case. A slight elongation of the overall shape of the $N$ = 1, 2 states is discernible along the [110] direction in $k$-space, corresponding to an elongation along  [1$\bar{1}$0] in real space, as discussed before. The next set of higher lying states $N$ = 3, 4 exhibit a node along [1$\bar{1}$0], while the states for $N$ = 5, 6 are oriented perpendicularly, along [110]. Note that the experimental resolution only allows us to map the 3, 4-states and the 5, 6-states together, hence the states 3,  4 and the states 5, 6 are simply doubled in Fig.~\ref{fig:mts}(a).

Figure~\ref{fig:mts}(b) shows our theoretical results for electrons in dot $D1$ (not elongated). We calculated the electron states for the elongated dot $D2$ as well and the results are qualitatively the same (not shown). 
The result of Fig.~\ref{fig:mts}(b) are in good agreement with the experimental findings. The theoretical and experimental results conclusively point to the filling ``Sequence I" of Fig.~\ref{fig:filling}(a). This implies that the $P_1$-$P_2$ splitting is large enough so that $P_1$ is fully occupied before $P_2$, following Aufbau. Had the $P_1$-$P_2$ splitting been smaller than the sum of exchange energy and the difference in Coulomb energies for one electron in $P_1$ and another in $P_2$, we would have obtained the charging sequence of Fig.~\ref{fig:filling}(b) \cite{bester03b} following the Hund rule that was observed in large dots\cite{tarucha96}. 
We choose to show the results of the dot with circular base $D1$ (while dot $D2$ is closer to the experimental situation) to emphasize the fact that shape elongation is not necessary to explain the results and to warn about the tempting conclusion that splitting of $P$-levels or wave function anisotropy is an indication of dot asymmetry.  
Our work reveals the importance of atomistic symmetry (not shape): Since our theoretical results from Fig.~\ref{fig:mts}(b) for dots with cylindrical base agree very well with our experimental findings, there is no need to assume shape asymmetry. Thus, the orientation of the $P$-states in Fig.~\ref{fig:mts}(b) is a result of the atomistic nature  of the underlying zincblende crystal lattice in contradiction with  
effective mass models that lead to degenerate and isotropic $P$-states.

\section{Results for Holes}

Figure~\ref{fig:mts}(c) shows the experimentally determined probabilities for the first six hole states. The data for $N=1,2$ shows again the shape of an $S$-state with, as was the case for the electrons, a slight geometric elongation along [110] in reciprocal space suggesting a slight elongation of the dot along the [1$\bar{1}$0] direction.
Close inspection of the $N=3$ and the $N=4$ hole states shows that they both exhibit a node along the [1$\bar{1}$0]-direction, whereas the $N=5$ hole state is almost circularly symmetric with a clearly developed minimum in the center and hole state $N=6$ exhibits nodes along [110]. 

For the theoretical calculations of Fig.~\ref{fig:mts}(d), we used the elongated dot $D2$ (see Table~\ref{tab:geometry}).

The transitions of the first two holes {\bf 0h $\rightarrow$ 1h, and   1h $\rightarrow$ 2h}  have no nodes and resemble the case of electrons. The general feature that the theoretical results show narrower, sharper peaks can be traced back to the fact that the experiment probes an ensemble of many quantum dots with slightly different shapes and hence slightly different transition energies, while the theory is performed assuming a single quantum dot. 

For the third and fourth holes the {\bf 2h $\rightarrow$ 3h, and   3h $\rightarrow$ 4h} transition amplitudes are anisotropic with peaks developing along the [1$\bar{1}$0] direction. This is the signature of the first hole $P$ state. Our work reveals the importance of inter-particle correlation effects: Indeed, a closer analysis of the theoretical results for the 2h$\rightarrow$3h transition shows that 91\% of the initial state is given  by the $h_0^2$-configuration (configuration as described in 
eq.~(\ref{eq:CI}), i.e, both holes occupy the first single-particle hole level $h_0$.  88\% of the final state is given by the $h_0^2h_1^1$-configuration where two holes are in state $h_0$ with $S$ orbital character and one hole is in $h_1$ with $P$ orbital character. This analysis shows that the tunneling hole is of orbital $P$-character. It is interesting to note how the mainly-single band electron and the multi-band hole $P$-states differ in their quasiparticle tunneling amplitude.

For the fifth hole, {\bf 4h $\rightarrow$ 5h}  is different and shows mostly isotropic features. This is the signature of the tunneling into the $D$ state. Indeed, 85\% of the initial state is given by the $h_0^2h_1^2$-configuration and 82\% of the final state by the $h_0^2h_1^2h_3^1$-configuration (where the last hole is in the $D$ state $h_3$). This shows, that the tunneling hole is mainly of orbital $D$-character. Comparison of the data in Fig.~\ref{fig:mts}(c) and (d) shows that even subtle differences in the shape of the wave functions can be resolved experimentally: The qualitative difference in the calculation between the 4h $\rightarrow$ 5h and the 5h $\rightarrow$ 6h transitions (namely equal or different amplitudes along the [110] and [1$\bar{1}$0] directions) are clearly reflected in the spectroscopic data. 
To emphasize the significance and clarity of the signature we obtain experimentally and theoretically, we artificially simulated the tunneling into a $P2$ state instead of a $D$ state in Fig.~\ref{fig:d_vs_p}. The left panel shows the artificial situation where we fixed the final configuration to a $h_0^2h_1^2h_2^1$-configuration, where the last hole is in the $P2$ state $h_2$, following the Aufbau principle. The right panel is the repetition of our actual result where the hole tunnels into a $D$ state. Both figures being very different, we conclude that the signature is a strong indication that, indeed, the $D$-state is filled before the $P2$ state. This is one of our main findings. 

Recent theoretical calculations \cite{climente05} showed, that even for a conventional filling of the shells, following the Aufbau principle, the magnetic field dependent charging may describe the experimental results of Reuter {\it et al.}\cite{reuter05} therefore challenging the interpretation of an unusual shell filling. We believe that our magnetotunneling spectroscopy results give additional evidence to supports the scenario of Reuter {\it et al.}\cite{reuter05} and He {\it et al.}\cite{he05} of a violation of the Aufbau principle.

For the sixth hole, {\bf 5h $\rightarrow$ 6h},  regains some anisotropic character with stronger maxima along the [110]-directions. The final state is still mainly given by  holes in $D$ states: $h_0^2h_1^2h_3^2$, but now to a somewhat lower percentage of 77\%. The remaining 23\% are configurations that include $P2$ state that have maxima along the [110]-direction. This is hence an effect of correlation that tend to become more important for heavily charged states. 
Our work reveals the importance of piezoelectricity on the shapes of wave functions: While it only weakly affects the energies \cite{bester06b} it modifies the single-particle wave functions significantly. Ignoring the piezoelectric effect leads to an anisotropic 4h$\rightarrow$5h transition and a more isotropic 5h$\rightarrow$6h transition, in contrast to the correct theoretical treatment and the experiment.

\section{Robustness of the Results}

To illustrate the robustness of our results we depict in Fig.~\ref{fig:mts_b250_h35} the theoretical results we obtain for hole charging on dot $D3$ that possesses a circular base. We believe that the slightly elongated dot $D2$ from Fig.~\ref{fig:mts}(d) fits better the experimental situation, as discussed in section~\ref{sec:Quantum_Dot_Shape}, but we want to illustrate the effect of shape on the results. The results in Fig.~\ref{fig:mts_b250_h35} are qualitatively very similar to Fig.~\ref{fig:mts}(d) for {\bf 0h $\rightarrow$ 1h},{\bf 1h $\rightarrow$ 2h},{\bf 2h $\rightarrow$ 3h} and {\bf 3h $\rightarrow$ 4h}. For the transition
{\bf 4h $\rightarrow$ 5h} the results are less isotropic in $D3$ than for the elongated dot $D2$ and shows more pronounced maxima along the  [1$\bar{1}$0] direction. The comparison of this transition with a tunneling of the hole into the second P-state (left panel Fig.~\ref{fig:d_vs_p}) show an even more pronounced difference than for the elongated dot $D2$: the orientation of the peaks are rotated by 90$^\circ$. Hence, the signature of tunneling into an orbital $D$-state is even stronger in a dot with circular base than for the elongated dot $D2$. However, the experimental results agree better with the calculations of dot $D2$ than $D3$, as we already concluded in section~\ref{sec:Quantum_Dot_Shape}. 
The results for {\bf 5h $\rightarrow$ 6h} are very similar for dot $D2$ and $D3$.

% Summary, Conclusions}

\section{Summary}

In conclusion, we measured and calculated the quasiparticle transition amplitudes resulting from a magnetotunneling experiment. Experiment and theory represent a novelty with the first experimental mapping of holes and the first atomistic calculation of quasiparticle transition amplitudes in quantum dots. We provide a direct evidence for the violation of the Aufbau principle for holes. The excellent agreement between the experiment and the theory allows us to further analyze and fully understand the nature of each of the many-body states probed. This analysis reveals the orbital character of the states, and shows the importance of the atomistic symmetry, correlations and piezoelectricity in the results. We believe that this type of approach delivers unprecedented insight into the electronic structure of self-assembled quantum dots and might be used for wave function engineering.

We acknowledge financial support of this work from the U.S. Department of Energy, Office of Science, Basic Energy Sciences, under Contract No. DE-AC36-99GO10337 (LAB-17) to NREL, from the DFG through the GRK384 and through the German Federal Ministry of Education and Research and under the ``nanoQUIT" Program.

\newpage

\begin{table}
\caption{Label and dimension, in nm, for the 3 dots considered. All dots are pure InAs with an overall lens shape. For dot $D2$ the base is elliptical while dots $D1$ and $D3$ have a circular base. \label{tab:geometry}}
\vskip 0.2cm
\begin{tabular}{c|ccc}
\hline
Dot & base along [1$\bar{1}$0] & base along [110] & height \\ \hline 
$D1$ & 25.0 & 25.0 & 5.0 \\ 
$D2$ & 26.0 & 24.0 & 3.5 \\
$D3$ & 25.0 & 25.0 & 3.5 \\ \hline
\end{tabular}
\end{table}

\begin{table}
\caption{Addition energies for electrons and holes in meV. 
The experimental values of electron and hole addition energies are extracted from 
Refs.~\onlinecite{miller97} and \onlinecite{reuter05} respectively.
``Theory $D1$'' are the results for an InAs lens-shaped dot with circular base of 25 nm diameter and 5 nm height. ``Theory $D2$'' is for an InAs lens-shaped dot with an ellipsoidal base of 26 nm (along[1$\bar{1}$0]]) x 24 nm (along [110])  and 3.5 nm height. ``Theory $D3$'' is for an InAs lens-shaped dot with a circular base of 25 nm diameter and 3.5 nm height.}
\label{tab:addition}
\vskip 0.2cm
\begin{tabular}{l|ccccc}
\hline
Addition &\multicolumn{5}{c}{Electrons}  \\
Energy   &$\Delta(1,2)$ & $\Delta(2,3)$ & $\Delta(3,4)$ & $\Delta(4,5)$ & $\Delta(5,6)$ \\ 
          \hline
Experiment   & 22 & 57 & 11 & 21 & 12  \\
Theory $D1$  & 22 & 63 & 19 & 22 & 19  \\
Theory $D2$  & 20 & 64 & 17 & 15 & 17  \\
Theory $D3$  & 21 & 66 & 16 & 18 & 16  \\ \hline
           & \multicolumn{5}{c}{Holes} \\ \hline
Experiment   &  24 & 34 & 17 & 23 & 15 \\
Theory $D1$  &  18 & 21 & 16 & 21 & 14 \\
Theory $D2$  &  11 & 15 & 13 & 15 & 13 \\
Theory $D3$  &  19 & 23 & 16 & 22 & 14 \\
\hline
\end{tabular}
\end{table}
\begin{figure*}
\caption{{\bf Figure can be found at http://www.sst.nrel.gov/nano\_pub/mts\_preprint.pdf}(Color online) Experimental [(a) (c)] and theoretical [(b) (d)] quasi-particle probability densities for electrons [(a) (b)] and holes [(c) (d)]. The directions given by arrows and labeled as [110] and [1-10], are real-space crystallographic directions (while the plots are in k-space). The Calculations are performed on a single quantum dot while the experiment probes an ensemble of quantum dots. For the electrons (holes), the reciprocal lattice vectors span a range of -6 to +6 $10^8$ m$^{-1}$ (-7 to +7 $10^8$ m$^{-1}$).}
\label{fig:mts}
\end{figure*}

\begin{figure}
\caption{{\bf Figure can be found at http://www.sst.nrel.gov/nano\_pub/mts\_preprint.pdf}Single-particle orbital filling sequence. Two possible scenarios for electrons are given in [(a) (b)] and our results for holes in [(c)].}
\label{fig:filling}
\end{figure}

\begin{figure}
\caption{{\bf Figure can be found at http://www.sst.nrel.gov/nano\_pub/mts\_preprint.pdf}(Color online) Contour plot of the transition probability for the transition 0h $\rightarrow$ 1h. The directions are given in real space while the plots are in reciprocal space, i.e., dot $D2$ is elongated along the [1$\bar{1}0$]-direction and so is the transition probability map. The reciprocal lattice vectors span a range of -5.4 to +5.4 $10^8$ m$^{-1}$ for a) and of -4.3 to +4.3 $10^8$ m$^{-1}$ for b) and c).}
\label{fig:elongation}
\end{figure}

\begin{figure}
\caption{{\bf Figure can be found at http://www.sst.nrel.gov/nano\_pub/mts\_preprint.pdf}(Color Online) (Left) Quasi-particle probability densities for the transition 4h $\rightarrow$  5h where the tunneling hole has been forced to occupy the $P2$ state. (Right) Result of the calculation where the hole tunnels into a $D$-state.}
\label{fig:d_vs_p}
\end{figure}

\begin{figure*}
\caption{{\bf Figure can be found at http://www.sst.nrel.gov/nano\_pub/mts\_preprint.pdf}(Color online) Calculated quasi-particle probability densities for holes in dot D3 with a circular base. The reciprocal lattice vectors span a range of -7 to +7 $10^8$ m$^{-1}$.}
\label{fig:mts_b250_h35}
\end{figure*}

\end {document}